\documentstyle[prl,aps,tighten,multicol]{revtex}

\begin{document}
\title{Escape time in anomalous diffusive media} 
\author{E. K. Lenzi$^1$\thanks{\rm e-mail: eklenzi@cbpf.br }, 
        C. Anteneodo$^2$\thanks{To whom correspondence should be 
        addressed, {\rm e-mail: celia@cbpf.br}} and 
        L. Borland$^1$\thanks{ {\rm e-mail: lisa@sphinx.com}} }

\address{$^1$
         Centro Brasileiro de Pesquisas F\'{\i}sicas, 
         R. Dr. Xavier Sigaud 150, \\
         22290-180, Rio de Janeiro, Brazil  \\
	 $^2$
	 Instituto de Biof\'{\i}sica, Universidade Federal  
         do Rio de Janeiro, \\ 
         Cidade Universit\'aria, CCS, Bloco G, 21941-900, \\
         Rio de Janeiro, Brazil }

\maketitle

\begin{abstract}

We investigate the escape behavior of systems governed by the 
one-dimensional nonlinear diffusion equation 
$\partial_t \rho\,=\,\partial_x[\partial_x U\rho]\,+\,D\partial^2_x
\rho^\nu$, 
where the potential of the drift, $U(x)$, presents a double-well 
and $D, \nu$ are real parameters. 
For systems close to the steady state we obtain an analytical 
expression of the mean first passage time, yielding a generalization 
of Arrhenius law. Analytical 
predictions are in very good agreement with numerical experiments 
performed through integration of the associated Ito-Langevin equation.
For $\nu\neq 1$ important anomalies are detected in comparison to the
standard Brownian case. These results are compared to those 
obtained numerically for initial conditions far from the steady state. 

\end{abstract}


\begin{multicols}{2}

\narrowtext
%%%%%%%%%%%%%%%%%%%%%%%%%%%%%%%%%%%%%%%%%%%%%%%%%%%%%%%%%%%%%%%%%%%%%%%%
%
%
\section{Introduction}

The old problem of surmounting a potential barrier, known as 
Kramers' problem, is doubtlessly relevant in connection with many 
topics, in fields ranging from physics to finance.   
It is a key ingredient to understanding phase-transitions in complex 
systems, both in  and far-from thermal equilibrium. In particular, 
the quantity known as the escape time (or mean first passage time)   
from one stable state to another has found numerous applications in 
a variety of interesting and novel problems. 
For example, it plays a key role in stochastic 
resonance \cite{stochasticresonance}, in describing fluctuation-induced 
transport such as occurs in kink motion \cite{kinkmotion} and ratchets 
\cite{ratchets}. Even the extent of chaos in Hamiltonian systems has been 
shown to have connections with this quantity \cite{chaosinhamiltonian}. 
A nice collection of these and other stochastically 
driven processes can be found in Ref \cite{fluctuationsandorder}.

However, all of the above examples have been formulated within a 
standard Brownian framework, for which diffusion properties are normal. 
In this paper we look at the  problem of calculating the escape time 
for systems exhibiting anomalous diffusion of the correlated type 
(in contrast to Levy type diffusion, which we do not discuss here). 
An understanding of escape time properties in such systems could open 
the door for understanding new stochastically driven 
phenomena. To our knowledge there has yet been 
little work done along these lines, although we are aware of some studies 
relating the anomalous transport properties on a random comb to the 
distribution of mean first passage times \cite{randomcomb}.  

The systems we are interested in are such that the diffusion 
is dependent on the density of particles $\rho$, resulting in a  
diffusion coefficient which is proportional to a power ($\nu-1$) 
of $\rho$. Many physical systems are well-described by this class of 
processes. Let us mention, amongst other examples, 
percolation of gases through porous media 
($\nu\geq 2$)\cite{porousmedia}, thin saturated regions in 
porous media ($\nu = 2$) \cite{saturated}, gravitational spreading 
of thin liquid films ($\nu = 4$) \cite{thinfilms},
heat transfer by  Marshak waves ($\nu = 7$) \cite{marshak}, 
surface growth ($\nu=3$) \cite{surfacegrowth},  
spatial diffusion of biological populations 
($\nu\geq 2$)\cite{populations}, plasma flows ($\nu<1$)\cite{plasma}. 
Explicitly, these processes are ruled by an equation of the type 
known in the literature as {\em porous media equation}\cite{porous}

\begin{equation} \label{PME}
\partial_t \rho(x,t)\,=\,D\partial^2_x[\rho(x,t)]^\nu,
\end{equation}
where $x$ is a dimensionless coordinate representing a bond-length, 
angle or any other chemical or physical state variable, 
$t$ is the dimensionless time and $\nu D>0$. Rewritting the nonlinear 
term as $\partial_x(D\nu \rho^{\nu-1}\partial_x \rho)$, 
it becomes evident that the restriction $D\nu>0$ guarantees that 
the flux will be from more dense to less dense regions.

Since the non-linearity in $\rho$ is known to lead to anomalous diffusion 
if $\nu\neq 1$ (namely superdiffusion for $\nu<1$ and subdiffusion for 
$\nu>1$\cite{lisa,ct}, as $<x^2(t)>\propto t^\frac{2}{\nu+1}$) 
important anomalies are also  expected when crossing over a barrier 
is involved. Precisely, we want to unveil here 
how escape properties are altered when $\nu\neq 1$. 

The paper is organized as follows.  In Sec. II we present 
the systems of interest and discuss some of their general features.   
Because fluctuations are determined  by $\rho(x,t)$ for  $\nu\neq 1$, 
the escape behavior will depend on the initial condition $\rho(x,0)$.
Therefore we first consider systems in the vicinity of the steady state,
a condition which allows analytical treatment. Numerical and analytical 
results for this case are presented in Secs. III and IV, respectively. 
In Sec. V we study numerically the escape behavior of systems far from 
the steady state, comparing the results with the previous ones. 
Finally, section VI contains concluding remarks. 
%%%%%%%%%%%%%%%%%%%%%%%%%%%%%%%%%%%%%%%%%%%%%%%%%%%%%%%%%%%%%%%%%%%%%%%%
%
%
\section{The system}  

Let us consider a set of identical particles immersed into a 
thermal environment such as that described by the porous media 
equation (\ref{PME}). Under the influence of an external bistable 
potential $U(x)$,  introduced in order to probe the escape behavior,  
the density of particles evolves following the nonlinear 
Fokker-Planck (FP) equation:

\begin{equation} \label{FP}
\partial_t \rho(x,t)\,=\,
\partial_x[\partial_xU(x)\rho(x,t)]\,+\,
D\partial^2_x[\rho(x,t)]^\nu.   
\end{equation}
This class of equations has been the object of diverse previous 
studies\cite{lisa,ct,ds}.

The stationary solution of Eq.~(\ref{FP}) is 
\begin{equation} \label{SS}
\rho_s(x) = [1-(\nu-1)\beta V(x)]_+^{\frac{1}{\nu-1}}/Z, 
\end{equation}
where $[f]_+=max\{f,0\}$, $Z$ is a (positive) normalization constant, 
$\beta=Z^{\nu-1}/(\nu D)$ and $V(x)=U(x)-U_o$, with $U_o$ the absolute 
minimum of the potential. 
In the limit $\nu\rightarrow 1$ the standard linear Fokker-Planck 
equation is obtained. In such a case, the steady state characterized 
by the Boltzmann-Gibbs 
distribution $\rho_s(x)\sim \exp(-U(x)/D)$, is recovered.
However, for $\nu\neq 1$,  the stationary solutions of Eq. (\ref{FP}) 
have the form of the Maximum Tsallis Entropy probability distributions, 
as already discussed previously \cite{lisa,ct,ds}, 
even in the absence of external drift \cite{porous,jou}.   
It is worth recalling that phemonena such as full developed 
turbulence \cite{cb}, the hadronic transverse moment 
distribution in high energy scattering process   
${\mbox {e}}^+{\mbox {e}}^-\rightarrow {\mbox {hadrons}}$ \cite{ec},   
among others, have been satisfactorily described in terms 
of distributions similar to (\ref{SS}) instead of the 
canonical stationary one. 

Steady state solutions are illustrated in Fig. 1 for a quartic potential. 
Note that a cut-off condition (Tsallis cut-off), yielding regions with 
null probability, arises in the $\nu>1$ case (see Fig. 1b). 
For a quartic potential the condition $\nu>-3$ must hold so 
that the solutions can be normalized. However,  the free-particle 
case requires $\nu>-1$  so we restrict our discussion to this regime.

The nonlinearity in the diffusion term of Eq. (\ref{FP}) accounts 
for the fact that the environment presents some kind of disorder or  
long range correlations in space-time leading to diffusion anomalies. 
The expression  $\beta=Z^{\nu-1}/(\nu D)$ can be interpreted as a 
generalized Einstein relation for this scenario.  Note that  
in disordered or correlated systems such  as those discussed here, 
the standard Einstein relation is expected to be recovered
{\em in the absence of disorder} \cite{report}.
This corresponds to the case of $\nu = 1$ yielding the well-known 
result $D=1/\beta$. Also, as was shown in \cite{ct}, 
the time-dependent form of these Einstein relations can be used to 
demonstrate the anomalous scaling properties of these nonlinear 
diffusion systems. For the free particle one obtains 
$<x^2(t)> \propto 1/\beta(t) \propto Z^2(t) \propto t^\frac{2}{\nu+1}$. 

The Ito-Langevin (IL) counterpart of Eq.~(\ref{FP}) reads\cite{lisa}

\begin{equation} \label{IL}
\dot{x}\;=\;-\partial_xU(x)+\sqrt{|D|}\,[\rho(x,t)]^\frac{\nu-1}{2}\eta(t),
\end{equation}
where $\eta(t)$ is a delta correlated Gaussian noise with zero mean 
and variance 2. 
In the particular case $\nu=1$, the standard Langevin equation for 
constant noise is recovered. 
It is noteworthy that this is a phenomenological description, 
in which the microscopic trajectories are determined by the macroscopic 
quantity $\rho$ when $\nu\neq 1$. 
Physically, this represents a kind of statistical feedback. 
As with state-dependent noise, it is to be seen as the influence of the 
environment, which is otherwise not explicitly taken into account by the 
equations of motion.  As a particle evolves, it interacts with
the environment such that it reacts to the collective density of states
around it. We can think of the subdiffusive case as a kind of 
"attraction" to the other particles: Particles tend to stay close to the
other particles, fluctuating not far from them. Conversely, we can think
of super-diffusive cases as a kind of reaction to the sparseness:
If the particle is in a highly populated region then it is in a sense
confined by the other particles, and fluctuations are not so large, but 
as soon as it gets into less dense regions it does not experience this
confinement and fluctuations can get very large. 

%%%%%%%%%%%%%%%%%%%%%%%%%%%%%%%%%%%%%%%%%%%%%%%%%%%%%%%%%%%%%%%%%%%%%%%%
%
%
\section{Numerical results in the vicinity of the steady state}

For numerical experiments we chose as prototype of double-well potential
the 
quartic polynomial $V(x)=ax^4+bx^3+cx^2+d$. 
The coefficients  were chosen  as in  Fig.~1, 
for which $(x_L,x_O,x_R)=(0,1,3)$, with $x_L$, $x_O$ and $x_R$
corresponding to the 
bottom of the left-hand well, the top of the barrier and  the bottom of 
the right-hand well, respectively. 
We studied the escape behavior close to the steady state. 
That is, once a population of a large number of particles has already 
attained the steady state described by Eq.~(\ref{SS}), a probe was injected
at $x_L$. 
Then its trajectory was obtained by solving, following the numerical 
scheme in ref. \cite{risken}, the IL (\ref{IL}) for 
$\rho(x,t)=\rho_s(x)$, starting from $x(t=0)=x_L$. 
Typical trajectories are displayed in Fig.~2. 
For $\nu>1$, fluctuations are reduced and trajectories result confined to 
the region within the cut-off boundaries (see also Fig.~1b); moreover,   
when the diffusion constant $D$ is smaller than a critical value 
$D_c$ (here $D_c\simeq 0.17$  for $\nu=2$), the state space becomes  
disconnected and crossings become forbidden. 
For $\nu<1$, the amplitude of noise is enhanced in the regions of low 
density and the entire space tends to be populated. 

We measured the mean first passage time, i.e., 
the average time interval $T(x_L\rightarrow x)$ that a particle at 
$x_L$ takes to reach for the first time a given state $x>x_L$. 
In Fig.~3 we present plots of $T(x)\equiv T(x_L\rightarrow x)$ vs. $x$. 
For $\nu\geq 1$ (Fig.~3a), plateaux become evident as $D$ approaches  
$D_c$ indicating that most of the time is spent overcoming the barrier 
around $x_O$. 
On the other hand, for $\nu<1$ (Fig.~3b), the passage time is 
sensitive to the exact final state and there is not a well defined 
plateau, even in the small-$D$ regime. Moreover, as $D$ decreases, 
the curves collapse to a limiting one for states below $x_R$, 
but grow faster above $x_R$, diverging in the limit $D\rightarrow 0$. 
The escape behavior seems to be discontinuous at $D=0$. In fact, for $D=0$ 
there is no diffusion, however, for finite $D$ the particle is attracted 
towards the deepest valley at $x_R$ and becomes trapped within a typical 
time interval which is bounded from above.  
This effect can be understood having in mind that fluctuations 
depend on $D$ not only through the factor $\sqrt{|D|}$ but also 
by means of the density through a factor that, for $\nu<1$, becomes 
very large outside the neighborhood of the absolute minimum where 
particles tend to concentrate as $D\rightarrow 0$. 
In other words, the deterministic case is not recovered when 
$D\rightarrow 0$ since the effective diffusion coefficient $D\rho^{\nu-1}$ 
does not vanish in that limit due to the singularity at $\rho=0$. 

%%%%%%%%%%%%%%%%%%%%%%%%%%%%%%%%%%%%%%%%%%%%%%%%%%%%%%%%%%%%%%%%%%%%%%%%
%
%
\section{Analytical considerations}

Let us show that these results can be understood analytically.  
For a system in the vicinity of the steady state, we can consider the 
following approximation for Eq.~(\ref{FP})

\begin{equation} \label{FPaprox}
\partial_t \rho(x,t)\,\simeq\,
\partial_x[\partial_xU(x)\rho(x,t)]\,+\,
D\partial^2_x[\{\rho_s(x)\}^{\nu-1}\rho(x,t)].
\end{equation}
Once the FP equation is linear, the problem of escape from a  well 
can be treated directly, following the same lines as for homogeneous 
processes characterized by time independent drift and diffusion 
coefficients \cite{gardiner}. Basically an equation for the 
probability that the particle is still within a 
given interval of state space at time $t$ is found using the corresponding 
backward Fokker-Planck equation and solved under appropriate boundary
conditions. 
In this way, one finds that the mean first passage time 
$T(x_1\rightarrow x_2)$, for $x_1<x_2$, is given by

\begin{eqnarray}   \nonumber
T(x_1\rightarrow x_2) \,&=&\, |\nu|\beta 
\int_{x_1}^{x_2} [1-(\nu-1)\beta V(y)]_+^{\frac{|\nu|}{1-\nu}} dy 
\\ \label{Tnu}
\times&&
\int_{-\infty}^y [1-(\nu-1)\beta V(z)]_+^{\frac{\mu}{\nu-1}} dz, 
\end{eqnarray} 
where $\mu=1$ if $\nu>0$ and $\mu=1-2\nu$ if $\nu<0$.
Expression (\ref{Tnu}) reproduces numerical experiments with 
excellent agreement as illustrated in Fig.~3.  

In Fig.~4 we show $T\equiv T(x_R)\equiv T(x_L\rightarrow x_R)$ 
as a function of $1/D$ (full lines), for different values of $\nu>0$, 
as calculated from Eq.~(\ref{Tnu}). $T$ represents a measure of the escape 
time from the left to the right-hand well, even 
in the $\nu<1$ cases where plateaux are not well defined. 
In the range $\nu>1$, $T$ diverges at a value $D_c$, 
defined by the cut-off prescription, below which the 
right-hand well becomes inaccessible. 
In the $0<\nu<1$ case, $T$ saturates as $1/D$ increases. 
The hyperdiffusive regime $\nu<0$ (hence $D<0$), where spreading is faster 
than ballistic, demonstrates the same general features discussed for 
the region $0<\nu<1$ but $|D|$ must be considered instead of $D$. 
For any $\nu$ and small $1/|D|$ the escape time follows the power law 
$T\sim\beta^\frac{3}{4}\sim 1/|D|^\frac{3}{\nu+3}$.
 
If $x_1\simeq x_L$ and $x_2\simeq x_R$, then, it is possible to find an 
approximate expression for the escape time $T$ 
when $|D|$ (hence $1/\beta$) is sufficiently small, noting that the 
integrands in Eq.~(\ref{Tnu}) present sharp peaks at $x_O$ and $x_L$ 
respectively. 
In that case the integrals can be evaluated by a 
saddle-point approximation extending the integration limits to the 
whole space. Following this procedure we arrive at 

\begin{equation} \label{Taprox}
T\, \simeq\, 
\frac{2\pi}{\sqrt{\omega_L\omega_O}} \frac{2|\nu|}{|\nu|+\mu}
\left( \frac{1-(\nu-1)\beta V(x_O)}{1-(\nu-1)\beta V(x_L)}
\right)^\frac{|\nu|+\mu}{2(1-\nu)}, 
\end{equation} 
where $\omega_L$ and $\omega_O$ are the frequencies at the bottom of the 
left well and at the top of the barrier, respectively. 
Expression (\ref{Taprox}) is a generalization of the Arrhenius law, 
which, as expected, 
is recovered in the limit $\nu\rightarrow 1$. In fact, in that limit,  
$T\simeq (2\pi/\sqrt{\omega_L \omega_O})\exp(\Delta V/D)$, 
where $\Delta V\equiv V(x_O)-V(x_L)$ is the barrier height.  

For comparison, the approximation given by Eq.~(\ref{Taprox}) is also 
exhibited in Fig.~4 (dashed lines). 
The approximation is good for large $1/|D|$, as expected. 
It works better for $\nu>1$. 
Let us comment the main features revealed by this expression. 
When $\nu>1$, it foresees the divergence of $T$ at finite $D$.  
In fact, $D_c$ is obtained from $1/\beta_c\simeq (\nu-1)V(x_O)$. 
When $\nu<1$, saturation of $T$ for large $1/|D|$ 
is also predicted (unless $V(x_L)=0$) since $\beta$ is an unbounded 
increasing function of $1/|D|$. 
If $V(x_L)=0$, then Eq.~(\ref{Taprox}) indicates that 
$T$ diverges for vanishing $|D|$. In particular, if $0<\nu<1$, 
$T\sim \beta^\frac{\nu+1}{2(1-\nu)} \sim 1/D^\frac{1}{1-\nu}$  
and the deterministic limit is achieved. 
In the limit $\nu\rightarrow 1$ 
the exponential growth of $T$ with $1/D$ is always recovered.  

%%%%%%%%%%%%%%%%%%%%%%%%%%%%%%%%%%%%%%%%%%%%%%%%%%%%%%%%%%%%%%%%%%%%%%%%
%
%
\section{Numerical results far from the steady state}

The problem in the vicinity of the steady state actually corresponds to a 
linear one with a state dependent diffusion coefficient. 
However, it allows an analytical treatment which can be had in mind as 
a reference when studying more general cases. In order to test how the 
previous results compare to those of a more general situation, 
we also performed numerical studies of the escape properties far from the 
steady state. Particularly, we studied the case where particles are 
injected all at the same time at $x_L$. 
This instance requires simultaneous
integration of the FP equation,  in order to follow the evolution of 
$\rho(x,t)$ starting from $\rho(x,0)=\delta(x-x_L)$, 
together with integration 
of the IL equation (\ref{IL}), starting from $x(t=0)=x_L=0$. 
Now, the parameter $\nu$ must lie in the region  $\nu>0$ due to the
divergence in Eq. (\ref{FP}). 
An implicit finite-difference scheme with centered space differences was 
employed for numerical integration of the nonlinear FP
equation\cite{nlpde}. 
The time evolution of the density is illustrated in Fig. 5. 

The escape time $T$ as a function of $1/D$ (symbols) obtained 
for different values of $\nu$ was included in Fig. 4.
Let us compare this case to the precedent steady one.
For sufficiently large $D$, $T$ is not sensitively 
dependent on the initial distribution and Eq. (\ref{Tnu}) fits well 
to the numerical results for any $\nu>0$, following the power law 
$T\sim 1/D^\frac{3}{\nu+3}$ derived above. 
On the other hand, for small $D$, crossing times become closer to those 
of the standard case $\nu=1$ for any $\nu$. This can be understood as 
follows. 
For $\nu>1$, passage times are smaller than those given by Eq. (\ref{Tnu}) 
since, as the distribution evolves,  
there is an initial passage even between regions disconnected at the 
steady state (see Fig. 5(a)). However, our results 
suggest that the divergence of $T$ for a finite critical $D$, close to
$D_c$, still occurs. 
On the other hand, in the range $\nu<1$, crossing times are larger 
than those given by Eq. (\ref{Tnu}) since now the 
density of particles is initially 
unfavorable for surmounting the barrier (see Fig. 5(b)). 
Saturation is not observed and the escape time increases with $1/D$  
apparently following a power-law. It is worth noting that, 
as derived above, 
a power-law with exponent $1/(1-\nu)$ is the one expected if the   
average effective potential felt by crossing particles has 
the absolute minimum at $x_L$ which is consistent with the observed 
density evolution (see Fig. 5(b)).

%%%%%%%%%%%%%%%%%%%%%%%%%%%%%%%%%%%%%%%%%%%%%%%%%%%%%%%%%%%%%%%%%%%%%%%%
%
%
\section{Final remarks}

Summarizing, we have obtained the escape time for systems exhibiting
anomalous 
diffusion due to a stochastic nonlinear dependency on the density. 
For steady-state conditions, we obtain an analytical expression for 
the mean first passage time whose predictions are in excellent agreement 
with numerical results (Fig.~3). This analytical expression yields 
a generalization of Arrhenius law. 
A behavior quite different from that of the standard Brownian 
case $\nu=1$ is depicted. Under close to stationary conditions, 
two regimes are detected:  
In the region $\nu<1$ (superdiffusion), the escape time $T$ saturates for 
vanishing $D$, if $V(x_L)\neq 0$, and grows with $1/D$ following a
power-law, otherwise.  
In the region  $\nu>1$ (subdiffusion), $T$ diverges at $D_c$ (Fig.~4).
 
These results give hints on what should be expected in more 
general cases. 
For systems far from the steady state,  
$T$ grows with $1/D$ apparently following 
a power-law in the superdiffusive cases 
while $T$ diverges at finite $D$ in the subdiffusive ones. 

{\bf Acknowledgements:}
We want to thank R. O. Vallejos for enlightening discussions. 
We acknowledge Brazilian Agencies, CNPq and FAPERJ for
partial financial support.
%%%%%%%%%%%%%%%%%%%%%%%%%%%%%%%%%%%%%%%%%%%%%%%%%%%%%%%%%%%%%%%%%%%%%%%%
%
%
\section*{Captions for figures}

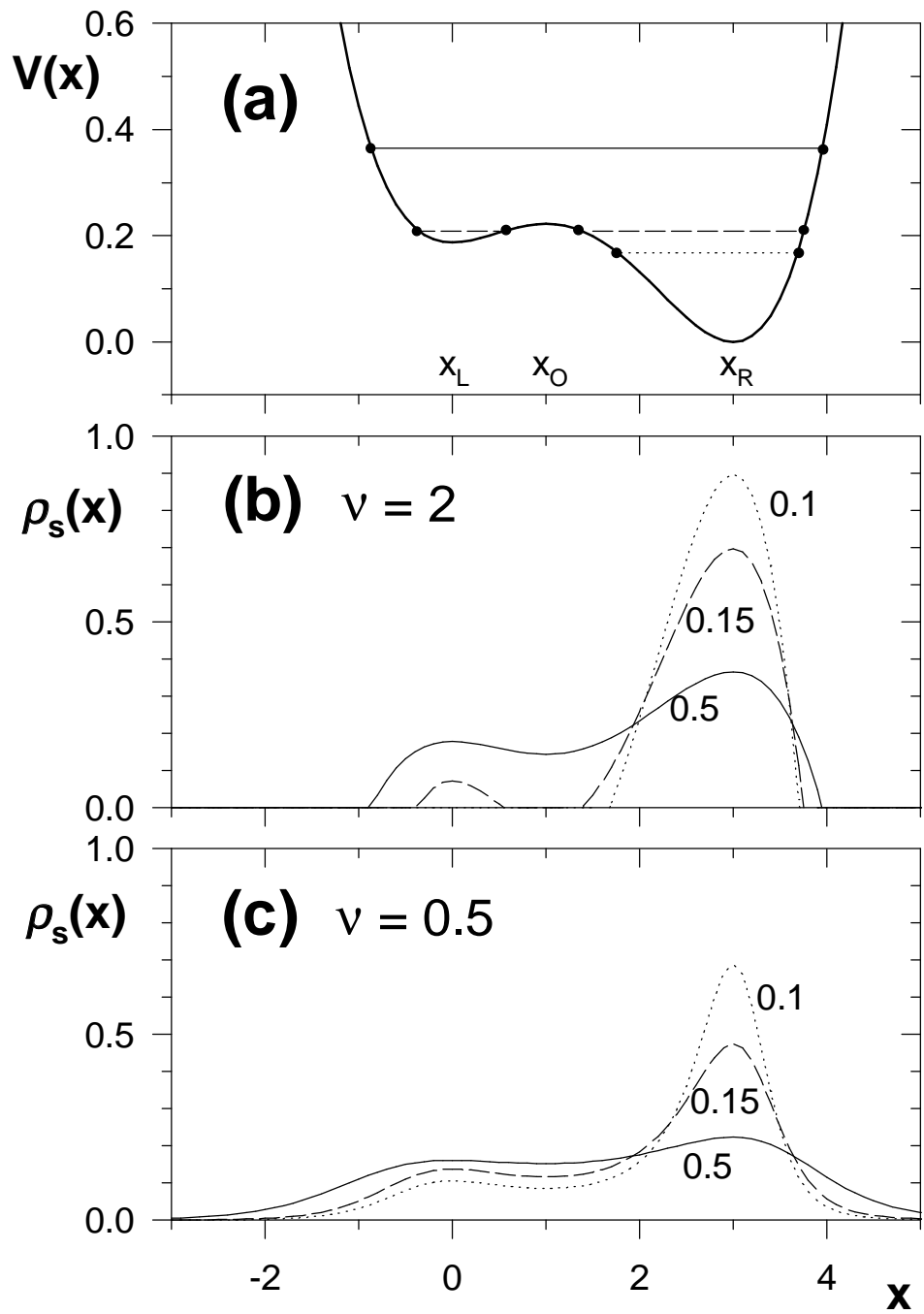
Figure 1: The cut-off condition. 
(a) Dimensionless double-well potential $V(x)= a x^4 + bx^3 + cx^2 + d$,
with $ a = 1/48,  b= -1/9, c = 1/8, d = 3/16$. 
The stationary distribution $\rho_s(x)$ is shown for  
$\nu=2$ (b) and $0.5$ (c), 
for different values of $D$ as indicated on the figure. 
For $\nu\leq 1$ the full state space is covered with power-law tails.
For $\nu>1$ a cut-off restricts the attainable  space. 
Observe in (b) that as $D$ decreases particles become more confined   
until only the neighborhood of the deepest valley is allowed. 
The horizontal lines in (a) represent the cut-off condition 
$V(x)=1/\beta$ which defines the allowed regions for $\nu=2$ and the 
same values of $D$ as in (b). All quantities are dimensionless.

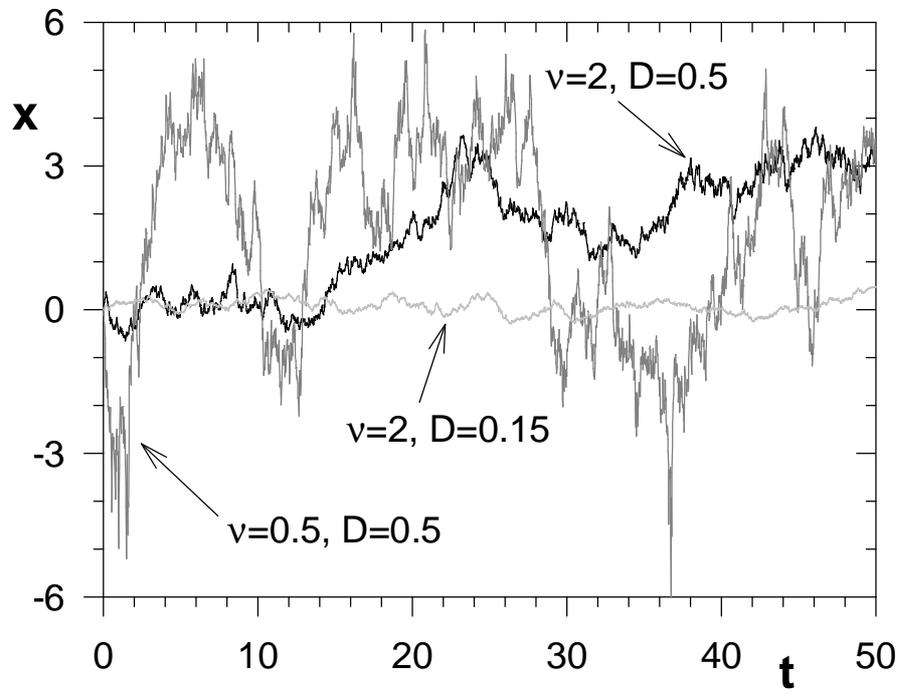
Figure 2: Typical trajectories $x$ vs. $t$ for $(\nu,D)=(0.5,0.5)$ 
(dark gray), $(2,0.5)$ (black) and (2,0.15) (light gray). 

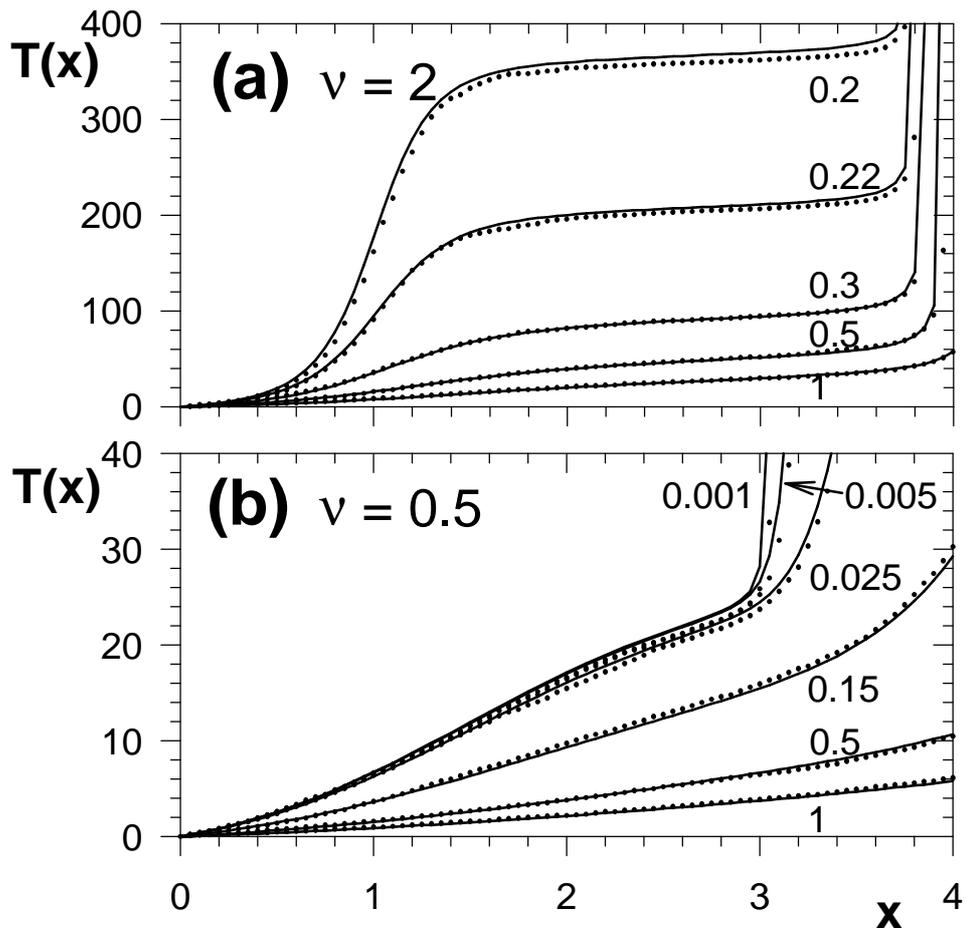
Figure 3: $T(x)\equiv T(x_L \rightarrow x)$ vs. $x$ for 
different values of $D$ 
indicated on the figure and $\nu=2$ (a), $0.5$ (b). 
Circles correspond to numerical experiments (mean value over 1000 
realizations) and full lines to theoretical prediction given by
Eq.~(\ref{Tnu}). 
 
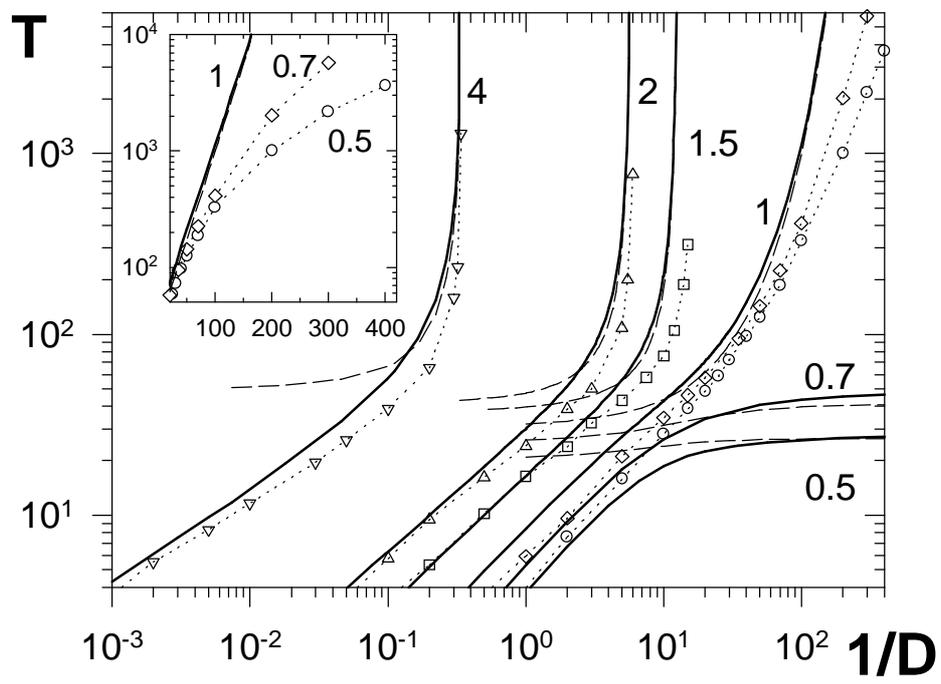
Figure 4: Escape time $T\equiv T(x_R)$ as 
a function of $1/D$, for different 
values of $\nu>0$ indicated on the figure. 
Full lines are generated from Eq.~(\ref{Tnu}). 
Dashed lines correspond to the low-$D$ approximation given by
Eq.~(\ref{Taprox}). 
Symbols correspond to the initial condition where all the particles 
(at least 1000) are injected at the same time at $x_L$. 
Dotted lines are guides for symbols. 
Insert: Detail (semi-log) of the low-$D$ region for $\nu\leq 1$.

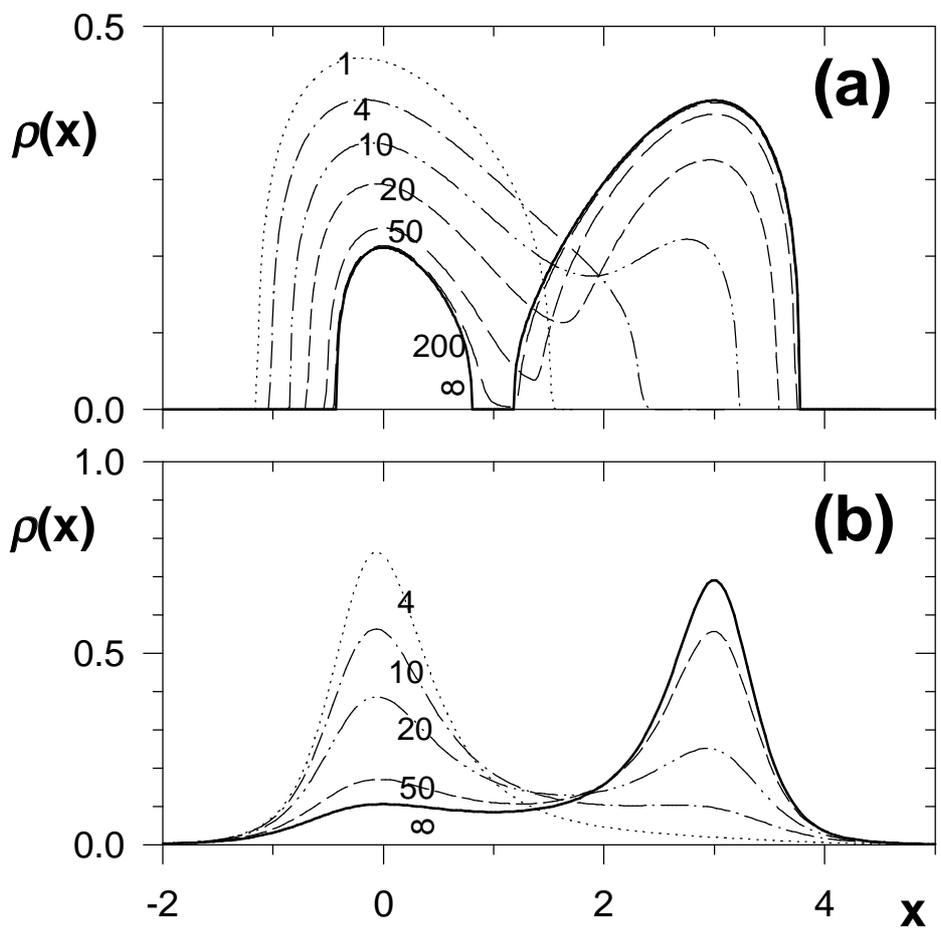
Figure 5: Time evolution of the density of particles    
obtained by numerical integration of Eq. (\ref{FP}) with
$\rho(x,0)=\delta(x)$ for 
($\nu$,$D$)=(4.0,2.5) (a) and (0.5,0.1) (b). 
The profiles correspond to times $t$ indicated on the figure.

\end{multicols}

\end{document}